\documentclass[twocolumn,aps,prl,showpacs,superscriptaddress,groupedaddress]{revtex4}
\usepackage[utf8]{inputenc}
\usepackage{float}
\usepackage{amsmath}
\usepackage{amssymb}
\usepackage{graphicx}
\usepackage{mathtools}
\usepackage{xcolor}
\usepackage{braket}
\usepackage{tikz}
\usepackage{changes}
\usepackage{cancel}
\usepackage{comment}
\makeatletter
\@ifundefined{textcolor}{}
{%
 \definecolor{BLACK}{gray}{0}
 \definecolor{WHITE}{gray}{1}
 \definecolor{RED}{rgb}{1,0,0}
 \definecolor{GREEN}{rgb}{0,1,0}
 \definecolor{BLUE}{rgb}{0,0,1}
 \definecolor{CYAN}{cmyk}{1,0,0,0}
 \definecolor{MAGENTA}{cmyk}{0,1,0,0}
 \definecolor{YELLOW}{cmyk}{0,0,1,0}
}

\usepackage{dcolumn}
\usepackage{bm}

\usepackage{bbm}

\makeatother

\global\long\def\da#1{#1^{\dagger}}

\global\long\def\L{\mathcal{L}}
\global\long\def\K{\mathcal{K}}

\global\long\def\ox{\mathcal{\otimes}}

\global\long\def\I{\mathbbm{1}}

\global\long\def\lin{Lindblad}
\newcommand*{\id}{\mathbbm{1}}
\global\long\def\be{\begin{equation}}
\global\long\def\ee{\end{equation}}
\global\long\def\linan{Lindbladian}

\begin{document}

\title{Anomalous Decay and Decoherence in Atomic Gases }

\affiliation{Department of Physics, Technion, Israel Institute of Technology, Haifa 32000, Israel}
\affiliation{RAFAEL, Science Center, Rafael Ltd., Haifa 31021, Israel}
\author{A.~Tsabary} \affiliation{Department of Physics, Technion, Israel Institute of Technology, Haifa 32000, Israel} \affiliation{RAFAEL, Science Center, Rafael Ltd., Haifa 31021, Israel}
\author{O.~Kenneth} \affiliation{Department of Physics, Technion, Israel Institute of Technology, Haifa 32000, Israel}
\author{J. E.~Avron} \affiliation{Department of Physics, Technion, Israel Institute of Technology, Haifa 32000, Israel}
\date{\today}
\begin{abstract}
Pair collisions in atomic gases lead to decoherence and decay.  Assuming that all {the} atoms in the gas are equally likely to collide  one is led to 
consider \linan\ of mean field type where the evolution in the limit of many atoms reduces to a {single qudit}
 Lindbladian with quadratic non-linearity. 
We  describe three smoking guns for {non-linear} evolutions: Power law decay and dephasing rates; {Dephasing}  rates that take a 
continuous range of values depending on the initial data and finally, {anomalous} flow of the Bloch ball towards  a hemisphere. 
 \end{abstract}

\maketitle

{Understanding and controlling decoherence is central to quantum sensing \cite{itano}, {time keeping} \cite{bolinger} and quantum computing \cite{nielsen}.}
A basic mechanism of decoherence and decay in atomic gases {is pair collisions} \cite{Fano1963,Budker2007,Allred2002,Bouchiat1966,stoof1988}:
 The  {internal} degrees of freedom of atomic vapors {decohere by exchanging}  energy and angular momentum   with  the spatial degrees of freedom \cite{happer,kominis-zeno}. 
{Such collisions can lead to non-linear evolutions \cite{happer1973}} {that deviate from the canonical Lindbladian evolution of linear open systems.  Nonlinear \linan\ {have been} treated  by linearization \cite{happer1973}, numerical calculation  and perturbation theory \cite{katz2013}.  {Here we shall describe} properties of collision-based nonlinear Lindblad evolutions using analytically solvable examples.}

  The kinetic degrees of freedom of the atoms are viewed as (parts of) 
a thermal bath.  
 We assume that all atoms are identical and any two atoms are equally likely to collide independent of how far they are. 
{The time scale of the problem is determined by 
 $\gamma> 0$,  the  (average) rate of collision of an atom with any one of the other $N$ atoms.} Since $N$, the number of atoms in the gas, is large, it is unlikely that a given pair will collide twice. {This is reflected in
$\gamma/N$, the rate of collision  of  any fixed pair,  being negligible as $N$ gets large.} 

Pair collisions are governed by a Poisson process, $\mathbf{N}(t)$, which counts the number of collisions up to time $t$. In a collision, the state of the pair changes 
by a  Kraus map $\K$ \cite{kraus,nielsen} which may be viewed as the generalization of the scattering matrix to open systems. 
The stochastic evolution equation for the pair, (in the interaction picture,) is governed by   
\be\label{e:ip}
\rho_2(t+dt)-\rho_2(t)= (\K-\id)\rho_2(t) \, d\mathbf{N}
\ee
Since $\mathbb{E}(d\mathbf{N}) =(\gamma/N)\, dt$, the  average evolution is 
\be\label{e:pair}
d\rho_2=\frac \gamma N(\K-\id) \rho_2\, dt, 
\ee
The factor $N^{-1}$  {is interpreted as}  rare events rather than weak interactions. 
Since $\K-\id$ has a \lin\ form \footnote{This follows easily by writing $\K\rho=\sum K_j\rho \da K_j$ with $\sum\da K_jK_j=\id$. } we shall henceforth denote it by $\L$. 

 It follows that the  density matrix of the internal degrees of freedom of the  gas evolves  by a \linan\ of  mean field type \cite{spohn,alicki1987}:
\begin{equation} 
\frac{d\rho^{(N)} }{dt} =\frac{\gamma}{N}\left(\sum_{k>  j=1}^{N}\L_{jk}\right)\rho^{(N)} . \label{e:mf}
\end{equation}
 $\L_{jk}$ are symmetric under interchange of atoms and we assumed $\L_{jj}=0$ (without loss of generality).

We  assume initial data {in which the atoms are uncorrelated}
\be\label{e:init}
\rho^{(N)}(0)=\rho_0^{\otimes N}
\ee
where $\rho_0$ may be a superposition of internal energy states. The evolution {takes place} in a Hilbert space whose dimension is  exponential in $N$.

It is a known fact about Eq.~(\ref{e:mf}) with initial data  Eq.~(\ref{e:init}), \cite{spohn,alicki1987}, 
that in the $N\to\infty$ limit,  the evolution preserves 
the product structure 
{of any finite subcluster}
and in particular {for any pair}:  
\be\label{e:partialtrace}
\rho^{(j,k)}(t)=Tr_{N-2}\rho^{(N)}(t)=\rho(t)\otimes\rho(t)
\ee
({For additional details on this equation see appendix \ref{babyalicki}}).
It follows  that the linear {$N$} body evolution in Eq.~(\ref{e:mf}), reduces in the large $N$ limit, to a non-linear evolution of a single atom  with quadratic 
non-linearity   \cite{spohn,alicki1987}
\begin{equation}
\frac{d \rho}{dt}= \gamma \,Tr_{{2}}\left(\L_{12}\rho\otimes \rho\right).\label{e:2nl}
\end{equation}
The evolution is trace and positivity preserving \footnote{Linear \linan s are also complete positivity preserving. However, the notion of complete positivity 
is not defined for non-linear systems.}.

{In the rest of this paper we shall consider examples of Kraus operators for which Eq.~(\ref{e:2nl}) can be solved exactly and where the solutions display unusual decoherence and decay properties.}   {Each of the Kraus maps we consider is associated with only one transition in the system. This {applies} when the other transitions are forbidden, or if the power spectrum of the bath is appropriate. The maps were chosen to demonstrate the richness {of} non-linear Lindblad equation.

For the sake of simplicity and geometric visualization, we  shall henceforth specialize to the case that $\rho$ describes a qubit,  
\be\label{e:Bvec}
\rho=\frac{\id +\mathbf{u}\cdot\boldsymbol{\sigma}}2,\quad  |\mathbf{u}|\le 1
\ee
The vector $\mathbf{u}$ represents the state  in the Bloch ball.
Under linear evolutions, the Bloch ball evolves into shrinking ellipsoids that eventually collapse on the stationary states \cite{algoet}. Quadratic evolutions allow for more complicated behavior, as we shall see.
\subsection{Polynomial decay}\label{s:dd}
 Consider the situation where excited pairs of atoms decay together to their ground states  through pairwise interaction. This scenario can be described by the \linan\ for a pair \footnote{This corresponds to weak collision in the sense that the Kraus operators for the collisions are $K_1= \sqrt {2 \epsilon} A$ and $K_2= \id -\epsilon \da A A$. We have suppressed $\epsilon$ in the \linan.}: 
  \be\label{e:decay}
\L_{12}\rho_2= A \rho_2 \da A-\frac 1 2 \{\da A A,\rho_2\},\ A=a_1\otimes a_2
\ee
with $a=\ket{0}\bra{1}$ the annihilation operator. Taking the partial trace in  Eq.~(\ref{e:2nl}) gives a quadratic \linan\ with an effective, state dependent, decay rate $\tilde\gamma(\rho)$
 \be\label{e:qdecay}
\frac{d\rho}{dt}=\frac{\tilde \gamma(\rho)}   2\,\left(2 a\rho \da a -\{\da a a,\rho\}\right), \ \tilde\gamma(\rho)=\gamma Tr \,(\da a a \rho)
 \ee 
 {Plugging the expression for $\rho$ Eq.~(\ref{e:Bvec}) in the Lindblad Eq.~(\ref{e:qdecay}) results in the} equation for the Bloch vector
 \be\label{e:3du}
\dot{\mathbf{u}}=
-\frac{\tilde\gamma} 2   (u_x,u_y,2 (u_z-1)), \quad \tilde\gamma=\gamma(1-u_z)/2
\ee
Since
\be
2{d\log u_x}=2{d\log u_y}={d\log (1-u_z)}=-{\tilde\gamma} \color{black} dt
\ee
the {trajectories} are parabolas, independent of  $\tilde\gamma(\rho)$, Fig.~\ref{Trajectory}.
The atoms eventually relax to the ground state which is the {(only)} stationary point.
%
The non-linearity only affects the schedule. 

To find the schedule, consider the equation for $u_z$, which decouples from the rest
\be\label{e:uz}
\dot u_z= \gamma(1-u_z)^2/2
\ee 
The solution is a $1/t$ decay law
\be\label{e:1/t}
1-u_z(t)=
\left(\frac 1 {1-u_{z}(0)\color{black}} +  \frac{\gamma t} 2 \right)^{-1}
\ee
The polynomial decay is a  smoking gun of non-linear evolution equation. 

\begin{figure}[h]

\begin{tikzpicture}[scale=1.5]
		\draw[ultra thick] (0,0) circle (1.05cm);
	
      \node[above] at (0,1) {$\ket{0}\bra{0}$}; 
            \node[below] at (0,-1) {$\ket{1}\bra{1}$}; 
              \node[right] at (1,0) {$\ket{+}\bra{+}$}; 
                 \node[left] at (-1,0) {$\ket{-}\bra{-}$}; 
         \draw[->,red,thick] (-0.0,-.95) -- (0,0.9); 
           \draw[red, line width = 0.50mm,<-]   plot[smooth,domain=0:1] (\x, {1-\x^2});
              \draw[red, line width = 0.50mm,->]   plot[smooth,domain=-1:0] (\x, 1-\x*\x);
    \end{tikzpicture}
\caption{\label{Trajectory}Trajectories of dephasing and decay in the
Bloch sphere.  The decay rate {has} $1/t$ {law}. }

\end{figure}
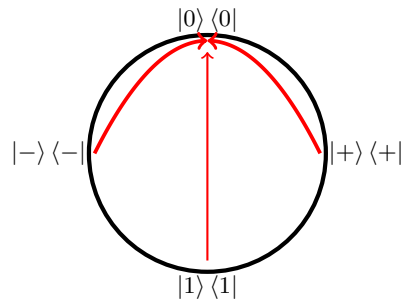

 The slow decay has a simple interpretation: 
  The effective rate $\tilde\gamma$  slows down as atoms  relax to the ground state because excited atoms find it harder to mate with a partner that would allow both to decay. 
 
\subsubsection*{{Master equation}}
Since the mean field limit, $N\to \infty$, need not commute with the long time limit, {let us} {show that the} slow decay in Eq.~(\ref{e:1/t}) is 
not an artifact of the mean-field approximation. 
 
We shall therefore solve  Eq.~(\ref{e:mf}) directly for the pair-\linan\ $\L_{j,k}$ given in Eq.~(\ref{e:decay}), without making the mean-field approximation, 
Eq.~(\ref{e:2nl}). We shall consider initial symmetric mixture of $P_0=\ket{0}\bra{0}$ and $P_1=\ket{1}\bra{1}$, but no superpositions.

Denote the symmetric  mixture of $2n$ qubits in $P_1$ and the rest in $P_0$,  by 
\begin{equation} 
R_{n}=\mathcal{S}\left(P_{1}^{\ox2n}\ox P_{0}^{\ox\left(N-2n\right)}\right),\ n\in\left\{ 0,\dots,\left\lfloor\frac{N}{2}\right\rfloor\right\}\label{Rn}
\end{equation} 
where $\mathcal{S}$  is the operator of symmetrization. For instance, for $N=3$
the symmetric states are, 
\begin{align}
\begin{array}{l}
R_{0}\end{array} & =P_{0}^{\ox3}\\
R_{1} & =\frac{1}{3}\left(P_{1}^{\ox2}\ox P_{0}+P_{1}\ox P_{0}\ox P_{1}+P_{0}\ox P_{1}^{\ox2}\right).\nonumber 
\end{align}
For simplicity, we henceforth assume that $N$ is even. 

The decay described by Eq.~(\ref{e:decay}) removes a pair of excited atoms and replaces them by a pair in the ground state:
\begin{equation}
\L_{\{12\}}\left(P_{b}\ox P_{c}\right)=\begin{cases}
P_{0}\ox P_{0}-P_{1}\ox P_{1} & b=c=1\\
0 & otherwise
\end{cases}.\label{L12 on Rn}
\end{equation}
It follows that  the evolution retains the form of a symmetric mixture
\begin{equation}
\rho_{N}\left(t\right)=\sum_{n=0}^{{N}/{2}}{p_n}\left(t\right)R_{n}.\label{cnRn}
\end{equation}
${p_n}\left(t\right)$ are probabilities. Note that the $1+N/2$ probabilities make an exponentially small fraction of the $2^N$ probability amplitudes of a {general} quantum state in the N-atom Hilbert space.

{Substituting Eqs.~(\ref{L12 on Rn}), ~(\ref{cnRn}) into the Lindblad Eq.~(\ref{e:mf}) yields}
\begin{align}
\frac{d}{dt}\rho_N(t) & =\frac{\gamma\color{black}}{N}\sum_{n=1}^{\frac{N}{2}}{p_n}\left(t\right)\left(\begin{array}{c}
2n\\
2
\end{array}\right)\left(R_{n-1}-R_{n}\right)
\end{align}
Equating the coefficients of $R_{n}$  gives the Master equation for the
vector of probabilities $\mathbf{p}=(p_{0}, \dots p_{N/2})$
\begin{equation}
\dot{\mathbf{p}}\left(t\right)=-\frac 1 N\,\mathbf{G}\, \mathbf{p}(t)\label{Gamma}
\end{equation}
where $\mathbf{G} $ is the upper triangular stochastic matrix \footnote{Columns sum to zero.} 
{           whose only non zero elements are}
\be   G_{nn}=-{G_{n-1,n}}=\gamma n(2n-1),\;\;   n=0,1,...N/2    \ee    
The solution of Eq.~(\ref{Gamma}) is simply $
\mathbf{p}\left(t\right)=e^{-\mathbf{G} t/N}\mathbf{p}\left(0\right)$. 
The rates  are given by the eigenvalues of  $\mathbf{G}/N $.  {Being} a triangular matrix,
the eigenvalues are given by the diagonal{, thus the eigenvalues of Eq.~(\ref{Gamma}), i.e. the decay rates, are $\gamma{n(2n-1)/N}$. The nonzero rates span from $\gamma/N$ to $\gamma(N-1)/2$, with a decreasing density as they increase}, see Fig. \ref{f:3}.

When the number of particles $N$ is very large, Eq.~(\ref{Gamma}) reduces to a first order PDE. To see this, let us introduce the continuous variable $x$ 
for the fraction of excited atoms:
\be
x=\frac{2n}{N}, \quad 0\leq x\leq1, 
\ee 
In the $N\to\infty$ limit, Eq.~(\ref{Gamma}) reduces  to the conservation law \cite{tsabari}.
\begin{equation}
\partial_t p= \partial_xj, \quad j=x^2 p
\label{cPDE}
\end{equation}
 The initial data is a probability distribution $p_0(x)$:
\be
p(x,t=0)=p_0(x)=\begin{cases}\ge 0 & x\in[0,1]\\
0 & $othewise$
\end{cases}
\ee
Eq.~(\ref{cPDE}) may be solved by the method of characteristics
\cite{haberman83}. 
\begin{figure}
\begin{center}

\begin{tikzpicture}[scale=2]
      \draw[->] (0,0) -- (3,0) node[right,thick] {$t$};
      \draw[->] (0,0) -- (0,1.2) node[above,thick] {$x$};
      \draw[scale=1,domain=0:3,smooth,ultra thick,variable=\t,blue] plot ({\t},{1/(1+\t});
      \node [left] at (0,1) {$x=1/(1+t)$}; 
         \draw[red,ultra thick] (0,0) -- (0,1) node[left] at (0,.55) {$p_0(x)$};
       \draw[red,ultra thick] (1,0) -- (1,1/2);
    \end{tikzpicture}

\caption{ The initial data $p_0(x)\ge0 $ on the interval $0\le x \le 1$ are marked by the red line at $t=0$.  $p_0(x)=0$  for $x\ge 1$. The initial data are dragged along the characteristics and hence stay below the  characteristics  emanating from $x=1,t=0$.  $p(x,t)=0$  above the blue hyperbola. }\label{f:char}
\end{center}
\end{figure}
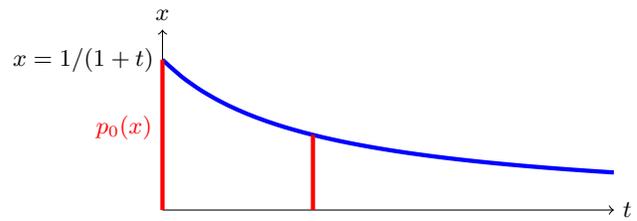
One finds {(see appendix \ref{PDEsol} for more details)}
\begin{equation}\label{e:final}
p\left(x,t\right)=(1-xt)^{-2} p_0\left(\frac{x}{1-xt}\right)
\end{equation}
Taking $p_0$ which is sharply localized one sees that the long time decay of the initial data toward the ground state has $1/t$ behavior, see Fig. \ref{f:char}.

The power {law} decay in the limit $N\to\infty$, due to a continuum of rates  with diverging density near zero, Eq. (\ref{e:final}), coincides with 
the power {law} decay in Eq.~(\ref{e:1/t}) which is due to nonlinearity  \footnote{Under the identification $2x=1-u_z$.} .

\begin{figure}[H]
\centering{}\includegraphics[scale=0.9]{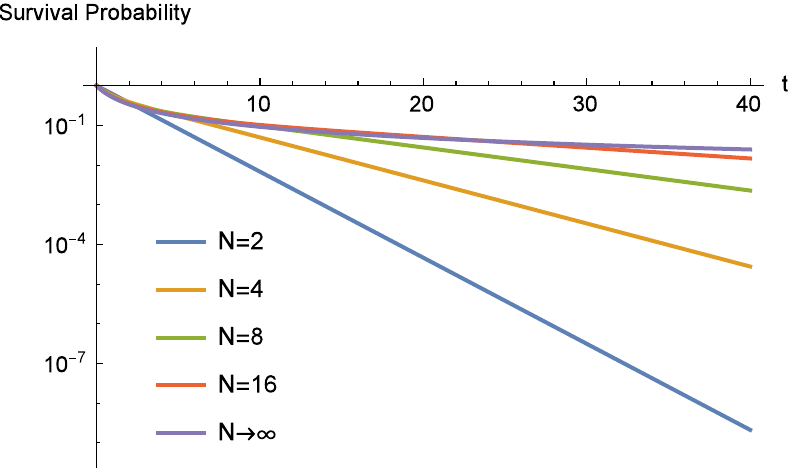}\caption{\label{P(N,t)}The survival probability of excited qubits as a function of time. 
The probabilities
are computed for an initially excited state that evolves according
to Eqs.~(\ref{Gamma}), {for various}  $N$.
The analytic solution for $N=\infty$, Eq.~(\ref{e:final}), is also plotted. The finite $N$ curves tend to the  $N=\infty$ curve as the number of particles increases and mutually agree for times smaller that $O(N)$, see \cite{tsabari}.}\label{f:3}
\end{figure}

\subsection{Continuum of dephasing  rates}
 Consider a situation where atoms do not  dephase  spontaneously, but do so in pairs when atoms collide {where the}   \linan\  for a colliding pair {is}: 
  \be\label{e:lede}
\L_{12}\rho_2=\gamma\big[K\otimes K,[K\otimes K,\rho\otimes\rho]\big] ,\quad \da K=K
\ee
Taking the partial trace as in Eq.~(\ref{e:2nl}) gives a quadratic \linan\ with an effective, state dependent, dephasing rate $\tilde\gamma(\rho)$ describing the gas of atoms
 \be\label{e:3d}
\frac{d\rho}{dt}=
\tilde\gamma(\rho)\, \big[K,[K,\rho]\big], \quad \tilde\gamma(\rho)=\gamma \,Tr\, (K^2 \rho) 
\ee

Choose the Pauli matrix $\sigma_z$ so that  $\sqrt 2 K=\id \cos\theta + \sigma_z\sin\theta$ \footnote{With  $\sin\theta\neq 0,\pi$ to avoid the trivial case.}. 
The equations of motion for the Bloch vector {that follow from Eq.~(\ref{e:3d})} are
 \be\label{e:3du2}
\dot{\mathbf{u}}=
-g (u_x,u_y,0), \ g= {\gamma } \sin^2\theta (1+u_z\sin 2\theta)
\ee 
The z-axis is the stationary manifold and $u_z(t)$ {is a constant of motion (}and hence also $g$ {)}. The orbits are radial {(in $x,y$)} with constant $u_z$ and the schedule is exponential
\be
{u_x(t)}={u_x(0)}e^{-g(u_z)t} 
\ee 
(See Fig. \ref{f:rates}). The rates $g(u_z)$ depend on the initial condition $u_z$ and take values in an interval
\be
g\in  {\gamma } \sin^2\theta [1-\sin 2\theta,1+\sin 2\theta]
\ee
The interval degenerates to a single point when $\theta=\pi/2$, which corresponds to the special case of linear evolution with  $\tilde\gamma=\gamma/2$.  An interval of decay rates is a smoking gun for the {non linear} evolution of the  qubits.

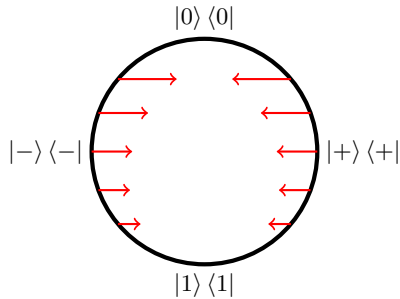
\begin{figure}[h!]
\begin{center}

\begin{tikzpicture}[scale=1.5]
		\draw[ultra thick] (0,0) circle (1cm);
		\foreach \x in {-40,-20,0,20,40}
      \draw[->,red,thick] (\x:1) -- ({cos(\x )-(\x+90)/ 250},{sin(\x )});
      \node[above] at (0,1) {$\ket{0}\bra{0}$}; 
            \node[below] at (0,-1) {$\ket{1}\bra{1}$}; 
              \node[right] at (1,0) {$\ket{+}\bra{+}$}; 
                 \node[left] at (-1,0) {$\ket{-}\bra{-}$}; 
                    	\foreach \x in {140,160,180,200,220}
      \draw[->,red,thick] (\x:1) -- ({cos(\x )-(\x-270)/ 250},{sin(\x )});
    \end{tikzpicture}

\caption{Dephasing with a continuum of rates. {The vectors represent} the exponential dephasing rate to the fixed point, $-\frac{d}{dt}\log \left|\mathbf{u}(t)-\mathbf{u}(\infty) \right|$.}\label{f:rates}

\end{center}
\end{figure}

\subsection{Flow to a Bloch hemisphere }\label{s:dds}
Consider a process where pairs of excited atoms in the singlet state $\sqrt{2} \ket{s}= \ket{01}-\ket{10}$ decay to their ground state $\ket{g}=\ket{00}$ by collisions. A simple Kraus operator that describes this process is 
\be
\K\rho =K_1\rho \da  K_1+ K_2\rho \da  K_2
\ee
with
\be
K_1= \ket{g}\bra{s}, \quad K_2= \id - \ket{s}\bra{s}
\ee
Substituting in  Eqs.~(\ref{e:pair},\ref{e:2nl}) gives
\be\label{e:ww}
\frac{d\rho}{dt}={\frac{\sigma_z} 4} \,\big(1- Tr(\rho^2)\big)={\frac{\sigma_z}2\det\rho}
\ee
 ({See appendix \ref{krausder} for more details}).
The corresponding equation for the Bloch vector is
\be
4\dot{\mathbf{u}}=  \left(0,0,  1-\mathbf{u}\cdot\mathbf{u}\right)
\label{bloch3}
\ee
$u_x$ and $u_y$ are constants of motion.  The decay of $u_z$ to the northern hemisphere is given by
\be\label{e:hemi}
 u_z(t)= g\  \tanh\left(\frac{g  t}4+\tanh^{-1}\frac{{u_z}(0)}g\right)
\ee
where $ g^2= 1-u_x^2-u_y^2$. All the points of the Bloch ball float up to the upper hemisphere with  a continuum of rates in the interval $ [0,1]$. 
 This is a third smoking gun for {non linear} evolution.

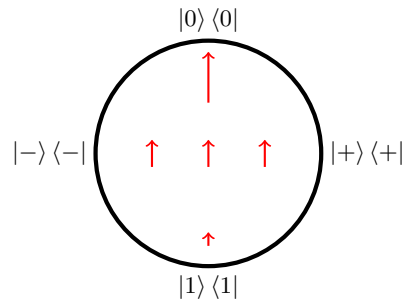
\begin{figure}[h]
\begin{tikzpicture}[scale=1.5]
		\draw[ultra thick] (0,0) circle (1cm);
	
      \node[above] at (0,1) {$\ket{0}\bra{0}$}; 
            \node[below] at (0,-1) {$\ket{1}\bra{1}$}; 
              \node[right] at (1,0) {$\ket{+}\bra{+}$}; 
                 \node[left] at (-1,0) {$\ket{-}\bra{-}$}; 
      \draw[->,red,thick] (0.5,.-.12) -- (.5,.12);
       \draw[->,red,thick] (-0.5,.-.12) -- (-.5,.12);
         \draw[->,red,thick] (-0.0,-.82) -- (0,-.7);
         \draw[<-,red,thick] (-0.0,.9) -- (0,.45);
          \draw[->,red,thick] (-0.0,-.12) -- (0,.12);
    \end{tikzpicture}
\caption{\label{f:pure}Trajectories of a purifying channel to the upper Bloch hemisphere. {The vectors represent the exponential flow rate to the fixed point.}}
\end{figure}

{\bf Summary:}
{We describe models of collisions where the mean field equations are solved  exactly and display anomalous behavior of decoherence and decay}:  Power law decay  to the ground state; A continuous interval of  dephasing rates that depend on the initial data; And finally, flow to the hemisphere of the Bloch ball.  None of  these features can occur in (time independent)  linear \linan\ of finite dimensional systems.

\section*{Acknowledgements}
The research is supported by the ISF {and Rafael LTD}. We thank Yonatan Rosmarin {f}or  a helpful observation.
\appendix
\section{Finite clusters remain uncorrelated}\label{babyalicki}

\maketitle
Our aim in this appendix is to  explain why Eq. \ref{e:2nl} holds. For rigorous proofs see \cite{spohn,alicki1987}.

Note that Eq.~(\ref{e:2nl}) immediately follows from Eq.~(\ref{e:partialtrace}), which states that the reduced 2-body density matrix
\[
\rho^{(12)}=Tr_{3,\dots N}\rho
\]
preserves the product structure when $N$ is large, i.e.
\[
\rho^{(12)}(t)=\rho^{(1)}(t)\otimes \rho^{(2)}(t)
\]
It is instructive to see first what lies behind the proof.  Consider
\begin{align*}
\dot\rho^{(12)}&=Tr_{3,\dots, N}\L\rho =\frac 1 N \L_{12}\rho^{(12)}+\\ & \frac 1 N\sum_{j>2} \left(Tr_j \L_{1j}\rho^{(12j)}
+ Tr_j \L_{2j}\rho^{(12j)}\right)
\end{align*}
The only term that describes direct interaction that can lead to buildup of the correlations in the 1-2 cluster, is the first term. This can be neglected when $N$ is large. The remaining terms, of $O(1)$, do not couple the pair 1-2. This is fundamentally why the product structure of the initial data in small clusters is preserved by the mean field evolution.    

Eq.~(\ref{e:mf}) gives rise {to an} infinite hierarchy of coupled equations for the partial traces 
\[
\dot\rho^{(1\dots k-1)}=\sum_{i}^{k-1} Tr_{k}\L_{ik}\rho^{(1\dots k)}+O(k^2/N)
\]  
The last term, describing  interactions within the k-cluster, can be neglected when $k^2\ll N$.
Assuming the limit $N\rightarrow\infty$ exists,
\footnote{The rigorous proofs, \cite{spohn,alicki1987}, establish the existence of the limit.}, gives the simpler hierarchy
\begin{align}\label{e:heir}
\dot\rho^{(1\dots k-1)}=\sum_{i}^{k-1} Tr_{k}\L_{ik}\rho^{(1\dots k)}
\end{align}
We shall now verify that Eq.~(\ref{e:heir}) is satified by the ansatz
\begin{align}\label{e:ans1}
\rho=\sigma^{\otimes N}
\end{align}
and obtain the equation for $\sigma$.

Substituting ansatz (\ref{e:ans1}) in Eq.~(\ref{e:heir}) gives
\begin{align}\label{e:subs}
&\sum_{i}^{k-1} \sigma^{\otimes i-1}\otimes\dot\sigma\otimes\sigma^{\otimes k-i-1}=\sum_{i}^{k-1} Tr_{k}\L_{ik}\sigma^{\otimes k}\\ &\nonumber=\sum_{i}^{k-1} \sigma^{\otimes i-1}\otimes Tr_{2}\left(\L_{12}\sigma\otimes\sigma\right)\otimes\sigma^{\otimes k-i-1}
\end{align}
It is evident that Eq.~(\ref{e:subs}) holds provided $\sigma$ satisfies the non-linear Lindblad equation
\begin{align*}
\frac{d \sigma}{dt}= Tr_{{2}}\left(\L_{12}\sigma\otimes \sigma\right)
\end{align*}
 which is  Eq.~(\ref{e:2nl}).
 
\section{Solution to the PDE for the probability distribution}\label{PDEsol}
Consider the partial differential equation for the probability distribution
for the fraction of excited atoms in the ensemble $p\left(x,t\right)$,
\begin{equation}
\frac{\partial}{\partial t}p=\frac{\partial}{\partial x}\left(x^{2}p\right)
\end{equation}
After some rearranging, 
\begin{equation}
2xp=\frac{\partial}{\partial t}p-x^{2}\frac{\partial}{\partial x}p\label{cPDE2}
\end{equation}

Eq. \eqref{cPDE} may be solved using the method of characteristics
\cite{haberman83}, which reduces solving the partial differential
equation to solving a set of ordinary differential equations. This
is done by introducing a curve in the $x,t$ plane along which the
partial differential equation transforms into an ordinary differential
equation,

\begin{equation}
\frac{d}{ds}p\left(x\left(s\right),t\left(s\right)\right)=F\left(p,x\left(s\right),t\left(s\right)\right),\label{cODE}
\end{equation}
where $s$ is a variable associated with the curve, and $F$ is a
some function of $p,x,t$.

The left hand side of the equation above may be rewritten as,

\begin{equation}
\frac{d}{ds}p\left(x\left(s\right),t\left(s\right)\right)=\frac{\partial p}{\partial t}\frac{dt}{ds}+\frac{\partial p}{\partial x}\frac{dx}{ds}.\label{chainrule}
\end{equation}

Equating the coefficients between Eqs. \eqref{chainrule} and \eqref{cPDE}
yields a set of three ODEs,

\begin{align}
\frac{dt}{ds} & =1\\
\frac{dx}{ds} & =-x^{2}\nonumber \\
F\left(p,x,t\right) & =2xp\nonumber 
\end{align}

The solution for $t\left(s\right)$ is 
\begin{equation}
t\left(s\right)=s
\end{equation}

The solution for $x\left(s\right)$is 
\begin{equation}
x\left(s\right)=\frac{1}{s+\frac{1}{x_{0}}}=\frac{1}{t+\frac{1}{x_{0}}}\Leftrightarrow x_{0}=\left(\frac{1}{x}-t\right)^{-1}.
\end{equation}

Finally, the solution for $p\left(s\right)$ is,

\begin{equation}
\frac{d}{ds}p  =2\frac{1}{s+\frac{1}{x_{0}}}p
\end{equation}
It follows that 
\begin{equation}
\ln p  =2\ln\left(s+\frac{1}{x_{0}}\right)+\ln\left(f\left(x_{0}\right)\right)=\ln\left(x^{-2}f\left(\frac{x}{1-xt}\right)\right)\nonumber 
\end{equation}

$f\left(x_{0}\right)$ is an integration constant. Therefore the solution
to the partial differential equation for $p\left(x,t\right)$ (Eq.
\eqref{cPDE}) is,

\begin{equation}
p\left(x,t\right)=x^{-2}f\left(\frac{x}{1-xt}\right).\label{c(x,t)}
\end{equation}

The function $f\left(\frac{x}{1-xt}\right)$ is determined by the
initial condition, 
\begin{align}
p\left(x,t=0\right) & =x^{-2}f\left(x\right)\equiv p_{0}\left(x\right)
\end{align}
substituting back in Eq. \eqref{c(x,t)} gives 
\begin{equation}
p\left(x,t\right)=\left(1-xt\right)^{-2}p_{0}\left(\frac{x}{1-xt}\right).
\end{equation}

\section{Derivation of Eq.~(\ref{e:ww}) from {the} Kraus map}\label{krausder}
Consider the Kraus map 
\begin{equation}
{\cal K}\rho=K_{1}\rho K_{1}^{\dagger}+K_{2}\rho K_{2}^{\dagger}
\end{equation}
with 
\begin{equation}
K_{1}=\left|g\right\rangle \left\langle s\right|\,,\qquad K_{2}=\I-\left|s\right\rangle \left\langle s\right|
\end{equation}
 where 
\begin{equation}
\ket g=\ket{00}\,,\qquad\ket s=\frac{\ket{01}-\ket{10}}{\sqrt{2}}.
\end{equation}
The mean field Lindbladian that arises from this Kraus map takes the
form 
\begin{align}
\L\rho & =Tr_{2}\left[\left({\cal K}-\I\right)\left(\rho\ox\rho\right)\right]\label{KtoL}\\
 & =\bra s\rho\ox\rho\ket s
\left(   Tr_2\left(\ket g \bra g+\ket s\bra s\right)  \right) 
\label{term1}\\
 & -Tr_{2}\left\{ \ket s\bra s,\rho\ox\rho\right\} .\label{term2}
\end{align}
 {It is convenient to use}
  the notation $\ket a=\sum_{j,k}a_{jk}\ket{jk}\:,\:\ket b=\sum_{j,k}b_{jk}\ket{jk}$.
 This gives  
\begin{align}
Tr_{2}\left(\ket a\bra b\rho\ox\sigma\right) & =a\sigma^{t}b^{t}\rho\\
Tr_{2}\left(\rho\ox\sigma\ket a\bra b\right) & =\rho a\sigma^{t}b^{t}
\end{align}
{In particular since  $\sqrt{2}\ket s=\sum_{jk}\varepsilon_{jk}\ket{jk}$,}
expression \eqref{term2} gives 
\begin{equation}
Tr_{2}\left\{ \ket s\bra s,\rho\ox\rho\right\} =\frac{1}{2}\left(\varepsilon\rho^{t}\varepsilon^{t}\rho+\rho\varepsilon\rho^{t}\varepsilon^{t}\right).
\end{equation}
One can validate that 
\begin{equation}
\varepsilon\rho^{t}\varepsilon^{t}\rho=\rho\varepsilon\rho^{t}\varepsilon^{t}=\I\det\rho,
\end{equation}
thus expression \eqref{term2} simplifies to 
\begin{equation}
Tr_{2}\left\{ \ket s\bra s,\rho\ox\rho\right\} =\I\det\rho.
\end{equation}
Expression \eqref{term1} is similarly simplified, 
\begin{align}
 & Tr\left(\ket s\bra s\rho\ox\rho\right)
\left(Tr_{2}\left(\ket g\bra g\right)+Tr_{2}\left(\ket s\bra s\right)\right)\\
 & =Tr\left(\frac{\det\rho}{2}\I\right)\left(\ket 0\bra 0+\frac{1}{2}\I\right)\nonumber \\
 & =\det\rho\left(\ket 0\bra 0+\frac{1}{2}\I\right)\nonumber 
\end{align}
where we used the identity: $Tr_{2}\left(\ket s\bra s\right)=\frac{1}{2}\I$.

The resulting Lindbladian is then 
\begin{equation}
\L\rho=\left(\det\rho\right) \left(\ket 0\bra 0+\frac{1}{2}\I-\I\right)=\frac{\det\rho}{2}\sigma_{z}.
\end{equation}
A density matrix of a qubit satisfies 
\begin{equation}
\det\rho=\frac{1-Tr\left(\rho^{2}\right)}{2},
\end{equation}
so the Lindbladian can be expressed as 
\begin{equation}
\L\rho=\frac{1}{4}\left(1-Tr\rho^{2}\right)\sigma_{z}.
\end{equation}

\end{document}